\documentstyle[preprint,aps,prd]{revtex}

\newcommand{\be}{\begin{equation}}
\newcommand{\ee}{\end{equation}}
\newcommand{\sst}{\scriptscriptstyle}

\newcommand{\ist}{\stackrel{\rm def}{=}}
\newcommand{\bea}{\begin{eqnarray}}
\newcommand{\eea}{\end{eqnarray}}
\def\r0{r_{\sst 0}}

\begin{document}

\title {Running Coupling Constants, Newtonian Potential and non
localities in the Effective Action}

\author{Diego A.\ R.\ Dalvit}

\address{{\it
Departamento de F\'\i sica, Facultad de Ciencias Exactas y Naturales\\
Universidad de Buenos Aires- Ciudad Universitaria, Pabell\' on I\\
1428 Buenos Aires, Argentina}}

\author{Francisco D.\ Mazzitelli}

\address{{\it
Departamento de F\'\i sica, Facultad de Ciencias Exactas y Naturales\\
Universidad de Buenos Aires- Ciudad Universitaria, Pabell\' on I\\
1428 Buenos Aires, Argentina\\
and\\
Instituto de Astronom\'\i a y F\'\i sica del Espacio\\
Casilla de Correo 67 - Sucursal 28\\
1428 Buenos Aires, Argentina}}

\maketitle

\begin{abstract}
We consider a quantum scalar field on an arbitrary gravitational
background. We obtain the effective {\it in-in} equations for
the gravitational fields using a covariant and non-local
approximation for the effective action  proposed by
Vilkovisky and collaborators.
{}From these equations, we compute the quantum corrections to
the Newtonian potential. We find logarithmic corrections which we
identify as the running of the gravitational constants.
This running coincides with the renormalization group prediction
only for minimal and conformal coupling.
\end{abstract}
\newpage

\section{INTRODUCTION}

The Effective Action (EA) is an useful tool for analyzing the quantum
corrections to the classical dynamics in quantum field theory. In
particular, the effective equations
derived from it should be the starting
point to investigate many interesting problems like the
influence  of quantum matter fields on the behaviour
of gravitational fields, both in cosmology and black hole physics.

The EA is a very complicated object, even in the one-loop approximation,
and it is necessary to develope approximation techniques in order
to evaluate it. A widely used approximation is the Schwinger DeWitt
expansion (SDW)\cite{sdwe}, which consists basically in an
expansion in derivatives
of the background fields. This expansion is useful in situations where
the background fields are slowly varying
with respect to the mass of the quantum fluctuations and
for the
analysis of the renormalizability of the theory.
However, many interesting physical effects are lost in this
approximation.
Alternatively, one can consider a situation where the background
fields are weak but rapidly varying.
In this case, it is possible to expand the EA in powers
of the curvatures of the background fields. The resulting
expansion has been recently investigated by Vilkovisky and
collaborators\cite{vilk1}-\cite{avra}, and it is in general
a {\it non-local} object.

On the other hand, there is a simple an intuitive way of taking
into account, at least partially, the quantum effects. In quantum
field theory, parameters like masses and coupling constants are
not constants but scale-dependent quantities. This is due
to vacuum polarization effects and the scale dependence is
dictated by the renormalization group equations. One can use this
fact to construct a `Wilsonian' effective action \cite{wilson},
which is basically the
classical action in which the parameters have been replaced by
their running counterparts. An  argument of
this type has been recently proposed to explain the dark matter
problem\cite{alamos}: due to quantum effects the Newtonian
potential should be
modified according to
\be
V(r)=-{G(\mu={1\over r})M\over r}
\ee
where $G(\mu)$ is the solution to the renormalization
group equations in a renormalizable theory of
gravity with $R^2$ - terms in the Lagrangian\cite{fradkin}.
As the theory is asymptotically free,
$G(r)$ is
an increasing function of $r$, and this may explain at least part
of the `missing' mass.  The running of $G$ may also induce
interesting cosmological and astrophysical effects\cite{perez}.

It is the aim of this work to analyze the relationship between
the non-local approach to the EA proposed by Vilkovisky et al,
the renormalization group, and the `Wilsonian' effective action.
In Ref. \cite{parker}, it was shown that the existence of
non-local terms in the effective action is linked to the
short distance behaviour of the theory and to the
renormalization group. The analysis was done in a non-covariant
weak-field
approximation, at the level of the {\it in-out} effective action. Here
we will extend that analysis: we will
be using a covariant effective action, we will work at the
level of the {\it in-in} (see below) semiclassical equations of motion
and we will see explicitly the running behaviour of the
gravitational constants in the Newtonian potential.
For simplicity, we will consider
a toy model in which we  quantize a scalar field on a
classical gravitational field. We will not include the
graviton loop in our calculations.

We would like to stress that our interest here is not to look
for measurable corrections to the Newtonian potential. Indeed,
we know a priori that these corrections are extremely small
in the toy model considered.  What we are going to discuss is
how to derive from `first principles' (i.e. the EA) the
scale dependence of $G$ in the gravitational potential.

The paper is organized as follows.
As a warm up, in the next Section we will obtain the
running behaviour of
the
electric charge in QED starting from the non-local EA.
In Section III we will
consider a free quantum scalar field of mass $m$ on an arbitrary
gravitational background. We will obtain the non-local
version of the effective action in powers of $-\frac{m^2}{\Box}$,
and the non-local effective equations. In Section IV we
will compute the quantum corrections to the Newtonian potential
and compare the results with the ones obtained from the
`Wilsonian' approach. Section V contains the conclusions
of our work.

\section {QUANTUM ELECTRODYNAMICS}

Due to the vacuum polarization, the electrostatic interaction potential
between point charges is modified to\cite{itz}
\be
V_{int}(r)=   \frac{e^{2}(r)}{4 \pi} = \frac{e^2}{4 \pi r}
\left[ 1 + \frac{e^2}{6 \pi^2} \int_{1}^{\infty} du e^{-2 m r u}
(1+\frac{1}{2 u^2}) \frac{\sqrt{u^2-1}}{u^2} + O(e^4)\right]
\label{eq:pot}
\ee
In the short distance limit ($m r \ll1$) we have
\be
e(r) = e \left[ 1 - \frac{e^2}{12 \pi^2}
        \ln{\frac{r}{r_{\sst 0}}} + O(e^4) \right]
\label{eq:itzi}
\ee
where $r_{\sst 0}$ is defined by $-\ln{m r_{\sst 0}} =
2 \gamma + \frac{5}{3}$.

On the other
hand, the solution to the renormalization group equation gives the following
running for the electric charge:
\be
e(\mu) = e(\mu_{\sst 0}) \left[ 1-\frac{e^2(\mu_{\sst 0})}{12 \pi^2}
\ln{\frac{\mu_{\sst 0}}{\mu}} + O(e^4) \right]
\label{eq:eren}
\ee
As can be easily seen from Eq. (\ref{eq:itzi}), in the
short distance limit the electrostatic interaction potential is just
the usual ${e^2\over 4\pi r}$ in which the electric charge
has been replaced by its running counterpart Eq. (\ref{eq:eren}),
with the additional rule that the mass scale $\mu$ is
replaced by the inverse of the distance $r$ and the mass scale
reference
$\mu_{\sst 0}$ is set at $\r0^{-1}$. Therefore the
`Wilsonian' argument gives the correct answer for the
electrostatic potential in the short distance limit.

We will derive here these old and well-known results using
the EA formalism, since this exercise will be a useful guide to
the more complex  calculation presented in Sections III and IV.
For simplicity we will quantize only the fermion
field ( with no classical component ), keeping the
electromagnetic field
as a classical background.

The classical action for QED in Euclidean space is given by
\be
S_{c} = {1\over 4}F_{\mu\nu}F^{\mu\nu} +
\bar{\psi} (\not\!\partial + i e \not\!\!A + i m) \psi
\ee
In the weak field approximation, the one-loop effective action obtained after
integrating out the fermions is given by
\be
S_{eff}= S_{c} + \frac{e^2}{2} \int d^{\sst 4}x d^{\sst 4}x'
A^{\mu}(x) \Pi_{\mu\nu}(x,x') A^{\nu}(x') + O(A^{\sst 4})
\ee
where $\Pi_{\mu\nu}$ is the usual vacuum polarization tensor for QED.
The renormalized effective action is
\be
S_{eff}= \frac{1}{4} \int d^{\sst 4}x F_{\mu\nu}
 \left[ 1 + \frac{e^2}{\pi^2} F(\Box) \right] F^{\mu\nu} +
 O(A^{\sst 4})
\label{sqed}
\ee
where
\be
F(\Box) = \frac{1}{8} \int_{0}^{1} (1-t^2)
  \ln \left[\frac{m^2 - \frac{1}{4} (1-t^2) \Box}{\mu^2} \right]
\label{fqed}
\ee
The modified Maxwell equations that derive
from Eq. (\ref{sqed}) are
\be
\left[1 + \frac{e^2}{\pi^2} F(\Box)\right] \partial_{\mu}F^{\mu\nu} =
J^{\nu}_{clas}
\label{euqed}
\ee
where we included a  classical source $J_{clas}^{\nu}$.

The form factor $F(\Box)$ admits the following integral
representation in terms of the massive Euclidean
propagator $(M^2-\Box)^{-1}$:
\begin{equation}
F(\Box)={1\over 8}\int_0^1 dt (1-t^2)\left[\ln
{(1-t^2)\over 4}+\int_0^{\infty}dz\left({1\over z+\mu^2}-
{1\over z+{4 m^2\over (1-t^2)}-\Box}\right)\right]
\label{rep}
\end{equation}
Therefore, we can regard $F(\Box)$ as a two point
function whose action on a test function $f(x)$ is given by
\begin{equation}
F(\Box) f(x)=\int d^4x'F(\Box)(x,x')f(x')
\end{equation}

All these equations
are valid in Euclidean space. To get the Minkowski version of them,
one should replace the Euclidean propagator by the Feynman one.
However, the equations thus obtained are neither real nor causal since
the effective action gives in-out matrix
elements instead of expectation values, making the
interpretation of the equations awkward. Alternatively, one can
use the Close Time Path (CTP) \cite{ctp} formalism to construct
an in-in effective action that produces real and
causal field equations for in-in expectation values \cite{variosctp}.

The CTP formalism involves a doubling of the degrees of freedom
and a generalization of the Feynman rules that includes both
the Feynman and Dyson propagators, as well as the two-point
Wightman functions (these functions carry the information about
the quantum state of the system). However, if one is interested
in the $in-in$ effective equations for the standard $in$-vacuum
state, this complication can be avoided.
Indeed,  in this situation it can be shown that
the $in-in$ version of the equations
is obtained by replacing  the Euclidean propagator by
the retarded one\cite{vilk2} in the integral representation of
the form factor Eq.\ref{rep}.
Alternatively, the $in-in$
form factor can be obtained in the $in-out$
formalism by taking twice the real and
causal part of the $in-out$ form factor\cite{jordan}.
We will denote the $in-in$
form factor thus obtained
by $F_{in}(\Box)$.

In particular, if the test function is time independent
\begin{equation}
F_{in}(\Box) f({\bf x})=F(\nabla^2) f({\bf x})=
\int d^3x'\int {d^3k\over (2\pi)^3}
e^{i{\bf k.(x-x')}} F(-{\bf k}^2) f({\bf x'})
\label{fourier}
\end{equation}
because the time integral of the retarded propagator
coincides with the Green function of the Laplacian.

In the short-distance limit $m^2 \ll -\Box$
the Euclidean field equation reduces to
\be
\left[ 1 - \frac{e^2}{12 \pi^2}
\ln (-\frac{\Box }{\mu^2}) \right]
\partial_{\mu}F^{\mu\nu} = J^{\nu}
\label{inout}
\ee

One often encounters the distribution
$G(-\frac{\Box}{\mu^2}) = \ln (-\frac{\Box}{\mu^2})$,
which will play a central role
in what follows. The action of the {\it in-in} counterpart of
$G(-\frac{\Box}{\mu^2})$ on a test function $f(x)$ is given by
(see Refs \cite{jordan} , \cite{horowitz} and
also \cite {jjg})
\begin{eqnarray}
G_{in}(-\frac{\Box }{\mu^2}) f(x) &=&
 \frac{2}{\pi} \int d^{\sst 4}{x'} \theta(x^{\sst 0}-{x^{\sst 0}}')
      \delta'((x-x')^2) f(x') \nonumber \\
                           &=&
 \frac{1}{2 \pi} \int_{0}^{\infty} du \int_{0}^{4 \pi} d\Omega
      \left[
 \ln (\mu u) \left. \frac{\partial f}{\partial u} \right|_{v=0}
         - \left. \frac{1}{2} \frac{\partial f}{\partial v} \right|_{v=0}
      \right]
\label{inin}
\end{eqnarray}
where $u$ and $v$ are respectively the standard retarded and advanced
coordinates with origin at the point $x$.
When the test function is time independent, Eq. (\ref{inin}) reduces
to
\begin{equation}
G_{in}(-{\Box\over\mu^2}) f({\bf x})=
G(-{\nabla^2\over\mu^2})f({\bf x})
\label{gin}
\end{equation}
as expected from Eq. (\ref{fourier}).

We are now ready to compute the modifications to the electrostatic
potential. Taking as a classical source a static point charge, the
modified Gauss law reads
\be
{\bf{\nabla} \cdot \bf{E}} -\frac{e^2}{12 \pi^2}
G(-\frac{\nabla^2}{\mu^2}) {\bf{\nabla} \cdot \bf{E}}  = e \delta^{3}(\bf{x})
\ee
The solution for the electric field is spherically symmetric
${\bf{E}} = E(r) {\bf{\hat{r}}}$ and we shall find it perturbatively
in powers of $e^2$
\be
\bf{E} = \bf{E^{\sst (0)}} + \bf{E^{\sst (1)}}
\ee
The leading order term is the classical contribution
\begin{eqnarray}
{\bf{\nabla}} \cdot {\bf{E^{\sst (0)}}} = e \delta^{3}({\bf{x}}) &
 \Longrightarrow & E^{\sst (0)} (r) = \frac{e}{4 \pi r^2}
\label{e0}
\end{eqnarray}
and the first quantum correction is given by
\be
{\bf{\nabla}} \cdot {\bf{E^{\sst (1)}}} =
        \frac{e^2}{12 \pi^2} G(-\frac{\nabla^2}{\mu^2})
{\bf{\nabla}} \cdot {\bf{E^{\sst (0)}}}
\ee
Therefore we have to evaluate the action of $G(-\frac{\nabla^2}
{\mu^2})$ on
the delta function. Using the Eqs. (\ref{fourier}) and (\ref{gin})
we readily obtain
\be
G(-\frac{\nabla^2}{\mu^2}) \delta^{3}({\bf{x}}) =
- \frac{1}{2 \pi r^3} -\ln{\mu^2} \delta^{3}({\bf{x}})
\label{logdelta}
\ee
where the last term gives a $\mu$-dependent correction to the
classical solution that will
be absorbed into the classical source. The quantum correction is
\be
E^{\sst (1)} (r) = E^{\sst (1)} (r_{\sst 0}) \frac{r_{\sst 0}^2}{r^2} -
    \frac{e^3}{24 \pi^3 r^2} \ln(\frac{r}{r_{\sst 0}})
\label{e1}
\ee
where $r_{\sst 0}$ is an arbitrary reference radius.
Integrating  Eqs. (\ref{e0}) and (\ref{e1})
and multiplying by the charge $e$
we get the electrostatic interaction
potential
\be
V_{int}(r) = \frac{e^2}{4 \pi r}
\left[
1-\frac{e^2}{6 \pi^2} \ln(\frac{r}{r_{\sst 0}}) + O(e^4)
\right]
\ee
{}From this equation the running behaviour of the electric charge follows, and
it coincides with that of Eq.(\ref{eq:itzi}).

\section{NON-LOCAL EFFECTIVE EQUATIONS FOR THE GRAVITATIONAL FIELD}

Let us now consider a quantum scalar field on a gravitational
background. The Euclidean action for the theory is
\be
S=S_{grav} + S_{matter}
\ee
where
\be
S_{grav}=-\int d^4x\sqrt g\left [{1\over 16\pi G}(R-2\Lambda)
+\alpha R^2 +\beta R_{\mu\nu}R^{\mu\nu}\right ]
\ee
and
\be
S_{matter}={1\over 2}\int d^4x\sqrt g [\partial_{\mu}\phi\partial^{\mu}\phi
+m^2\phi^2+\xi R\phi^2]
\ee
We have included terms quadratic in the curvature since in any
case they will appear in the renormalization procedure. As we will
use $\zeta$-function regularization, the constants $G,\Lambda ,\alpha$
and $\beta$ are finite (and dependent on a mass scale $\mu$).

The effective action for the classical gravitational field can be
obtained by integrating out the quantum scalar field. Formally
the result is
\be
S_{eff}=S_{grav}+{1\over 2}\ln\det \left[ {-\Box+m^2+\xi R\over\mu^2}
 \right ] \ist S_{grav}+\Gamma
\label{seff}
\ee
where $\mu$ is an arbitrary mass scale.

The evaluation of the above determinant in a general background is a
very complicated task.
Let us denote by  ${\cal{R}}$ either the Riemann tensor or any
of its contractions with
the metric.
When the gravitational field is slowly varying,
i.e. when $\nabla^n {\cal R}^m\ll m^{n+2m}$, one can use the
SDW
technique to get\cite{BD}

\bea
\Gamma &=& {1\over 32\pi^2}\int d^4x\sqrt g
           \left[
   {1\over 2} m^4 \ln ({m^2\over \mu^2})
   - m^2 a_1(x) \ln({m^2\over \mu^2})
           + \right. \nonumber \\
       & & \left. a_2(x) \ln({m^2\over \mu^2})
           +
   {1\over 2} \sum_{j\ge 3} a_j(x)(m^2)^{-j-4}(j-3)!
   \right]
\label{sdwe}
\eea
where we have omitted $\mu$-independent terms which redefine the classical
constants. The functions $a_j(x)$ are the coincidence limit of the
SDW coefficients, given by
\bea
a_0(x) & = & 1 \nonumber \\
a_1(x) & = & ({1\over 6}-\xi)R \\
a_2(x) & = & {1\over 180} R_{\mu\nu\rho\sigma} R^{\mu\nu\rho\sigma} -
     {1\over 180} R_{\mu\nu} R^{\mu\nu}	-
     {1\over 6} (\frac{1}{5} - \xi) \Box R +
             \frac{1}{2} \left( \frac{1}{6} - \xi \right)^2 R^2 \nonumber\\
       & \vdots &  \nonumber\\
a_n(x) &=& \nabla^{2n-2} {\cal{R}} + {\cal{R}} \nabla^{2n-4} {\cal{R}} +
           \cdots + \nabla \nabla {\cal{R}}^{n-1} + {\cal{R}}^{n}
\eea
The last line shows schematically the coincident limit of $a_n(x)$.

{}From the SDW expansion it is easy to derive the
scaling of the gravitational constants. The effective action
should not depend on the scale $\mu$. As a consequence,
taking $\mu$-derivatives in Eq. (\ref{seff}) we find
\bea
\mu{dG\over d\mu}&=&
 \frac{G^2 m^2}{\pi} \left(\xi -  \frac{1}{6}  \right)\label{rge1} \\
\mu{d\alpha\over d\mu}&=&
 -\frac{1}{32 \pi^2}
 \left[
 \left( \frac{1}{6} - \xi \right)^2 -\frac{1}{90}
 \right] \label{rge2}\\
\mu{d\beta\over d\mu}&=&
 -\frac{1}{960 \pi^2} \label{rge3}\\
\mu{d\over d\mu} \frac{\Lambda}{G}&=&
 \frac{m^4}{4 \pi}
\label{rge4}
\eea
which is the usual running for the gravitational constants.
We can use Eq.(\ref{rge1}) to construct a `Wilsonian' gravitational
potential. The scaling of $G$ is given by
\be
G(\mu)=G_0\left (1+{m^2G_0\over\pi}(\xi-{1\over 6})
\ln{\mu\over\mu_0}\right )
\label{Gmu}
\ee
so the Wilsonian potential is
\be
V(r)= -{G_0M\over r}\left (1-{m^2G_0\over\pi}(\xi-{1\over 6})
\ln{r\over r_0}\right )
\ee
In the next Section we will see if it is possible to derive
this potential from the
EA.

The SDW expansion is not useful for the analysis of the short distance
behaviour of the theory. As we have seen in the previous Section,
one should consider weak but rapidly varying background fields.
Assuming that $\nabla\nabla {\cal R}\gg {\cal R}^2$, one may try to sum up
in Eq. (\ref{sdwe}) all the terms which
contain a given power of the curvature.
This rather complicated calculation has been performed by Avramidy in Ref.
\cite{avra}. See also Refs. \cite{vilk2} for the massless case.
The result, up to second order in the curvature, is
\be
\Gamma=\Gamma_{local}+\Gamma_{non-loc}
\ee
where
\bea
\Gamma_{local}&=&{1\over 64\pi^2}\int d^4x\sqrt g \left[ m^4\left(
-{3\over 2}+\ln ({m^2\over\mu^2})\right) + 2 m^2\left(
-1+\ln ({m^2\over\mu^2})\right)(\xi-{1\over 6})R\right]\nonumber\\
\Gamma_{non-loc}&=&\frac{1}{32\pi^2}
\int d^4x\sqrt g\left[ RF_1(\Box)R+R_{\mu\nu}
F_2(\Box)R^{\mu\nu}+O(R^3)\right]
\label{ave}
\eea
and
\bea
F_1(\Box)&=&{1\over 2}\int_0^1 dt \left [ \xi^2-{1\over 2}\xi(1-t^2)
+{1\over 48}(3-6 t^2-t^4)\right ]\ln \left [{m^2-{1\over 4}(1-t^2)\Box
\over\mu^2}\right]\nonumber\\
F_2(\Box)&=&{1\over 12}\int_0^1  dt \,\, t^4
\ln \left [{m^2-{1\over 4}(1-t^2)\Box
\over\mu^2}\right]
\label{fgrav}
\eea
Note the similarities between these form factors and the corresponding
$F(\Box)$ that appears in QED (Eq. (\ref{fqed})).

{}From Eqs. (\ref{ave}) and (\ref{fgrav}), one can derive the effective
gravitational field
equations. As we are neglecting $O({\cal R}^3)$ terms in the effective action,
it makes no sense to retain $O({\cal R}^2)$ terms in the equations of motion.
Therefore, when doing the variation of the action, it is not necessary
to take into account the $g_{\mu\nu}$-dependence of the form factors.
Moreover, it is possible to commute the covariant derivatives acting
on a curvature, i.e., $\nabla_{\mu}\nabla_{\nu}{\cal R} =
\nabla_{\nu}\nabla_{\mu}{\cal R}
+O({\cal R}^2)$. After a straightforward calculation we find
\bea
& &\left[-{1\over 8\pi G}+{m^2\over 16\pi^2}(\xi -{1\over 6})(-1
+ \ln {m^2\over
\mu^2})\right]\left(R_{\mu\nu}-{1\over 2}Rg_{\mu\nu}\right)
\nonumber\\
& &- g_{\mu\nu}\left[{\Lambda\over 8\pi G}+
{m^4\over 64\pi^2}(-{3\over 2}+\ln {m^2\over\mu^2})\right] +
 \alpha H_{\mu\nu}^{(1)} +\beta H_{\mu\nu}^{(2)}=
{2\over\sqrt g}{\delta\Gamma_{non-loc}\over\delta g^{\mu\nu}}=
<T_{\mu\nu}>
\label{ecmov1}
\eea
where
\bea
H_{\mu\nu}^{(1)}&=& 4\nabla_{\mu}\nabla_{\nu}R - 4g_{\mu\nu}\Box R
+O(R^2)\nonumber\\
H_{\mu\nu}^{(2)}&=& 2\nabla_{\mu}\nabla_{\nu}R - g_{\mu\nu}\Box R
                    -2\Box R_{\mu\nu} +O(R^2)
\eea
and
\be
<T_{\mu\nu}>= \frac{1}{32 \pi^2} \left[
F_1(\Box) H_{\mu\nu}^{(1)} + F_2(\Box) H_{\mu\nu}^{(2)} \right]
\ee

Up to here we made no assumptions about the mass $m$.
In the large mass limit,
$m^2 {\cal R}\gg\nabla\nabla {\cal R}$ the SDW expansion
Eq.(\ref{sdwe}) is recovered
(up to $O({\cal R}^3)$). However, as in QED, we are interested
in the opposite
limit.
Let us assume that the typical scale of variation of the
gravitational field is much smaller than $m^{-1}$, that is,
$m^2{\cal R}\ll\nabla\nabla{\cal R}$. In this situation,
we can expand the functions
$F_1(\Box)$ and $F_2(\Box)$ in powers of $-\frac{m^2}{\Box}$.
The result is
\bea
F_1(\Box) &=&  \left[
               -{1\over {1800}} + \frac{5 \xi}{18} -\xi^2 +
       \frac{1}{2} \left(
       (\xi-\frac{1}{6})^2 - \frac{1}{90}
       \right)
       \ln ({- \frac{\Box}{\mu^2}})
               \right] +  \nonumber\\
          & &  \left[
       \frac{4}{18} - \xi + \xi^2 +
               \left( \xi^2 - \frac{1}{12} \right)
       \ln (-\frac{\Box}{m^2})
               \right]
               \left(-\frac{m^2}{\Box}\right)+ {\rm O} \left
         ( -\frac{m^2}{\Box} \right)^2
\label{f1grav}
\eea
and
\be
F_2(\Box) =  \left[ -\frac{23}{450} +\frac{1}{60} \ln(-\frac{\Box}{\mu^2})
               \right] +
               \left[-\frac{5}{18} +\frac{1}{6} \ln(-\frac{\Box}{m^2}) \right]
       \left( -\frac{m^2}{\Box} \right) +
               {\rm O} \left( -\frac{m^2}{\Box} \right)^2
\label{f2grav}
\ee
(It is possible to obtain exact expressions for $F_1$ and $F_2$
in terms of elementary functions. However, we will not need these
long expressions in what follows). Inserting the expansions
Eq.(\ref{f1grav}) and Eq. (\ref{f2grav}) into the effective equations
Eq.(\ref{ecmov1}) we get
\bea
& &\left[
    \alpha -
    \frac{1}{32 \pi^2} \left(
                       -{1\over {1800}} + \frac{5 \xi}{18} -\xi^2
                       \right) -
    \frac{1}{64 \pi^2} \left(
                  (\xi-\frac{1}{6})^2 - \frac{1}{90}
               \right) \ln ({- \frac{\Box}{\mu^2}})
\right] H_{\mu\nu}^{(1)} + \nonumber\\
& &\left[
    \beta -
    \frac{1}{32 \pi^2} \left(
               -\frac{23}{450} +\frac{1}{60} \ln(-\frac{\Box}{\mu^2}) \right)
\right] H_{\mu\nu}^{(2)} + \nonumber\\
& &\left[
   - {1\over 8\pi G}+{m^2\over 16\pi^2}(\xi -{1\over 6})(-1
+ \ln {m^2\over
    \mu^2})
\right]
\left(R_{\mu\nu}-{1\over 2}Rg_{\mu\nu}\right) + \nonumber\\
& & \left[
    \frac{m^2}{32 \pi^2}
      \left(
         \frac{4}{18}-\xi+\xi^2
         \right)
    \right] \frac{1}{\Box} H_{\mu\nu}^{(1)} -
    \left[
    \frac{5 m^2}{576 \pi^2}
    \right] \frac{1}{\Box} H_{\mu\nu}^{(2)} +\nonumber\\
& & \left[
    \frac{m^2 \xi^2}{32\pi^2}
    \right]
    \ln(-\frac{\Box}{m^2}) \frac{1}{\Box} H_{\mu\nu}^{(1)}
    -
    \left[
    \frac{m^2}{384 \pi^2}
    \right]
    \ln(-\frac{\Box}{m^2})\frac{1}{\Box}
    (H_{\mu\nu}^{(1)} - 2 H_{\mu\nu}^{(2)})
  = -T_{\mu\nu}^{clas}
\label{eqmov2}
\eea
where we have set the scale $\mu$ so that the cosmological
constant is zero and we included a classical source
$T_{\mu\nu}^{clas}$.

As with the SDW expansion Eq. (\ref{sdwe}), one can easily derive
$\mu$-dependence of the gravitational constants
from these modified Einstein equations.
Alternatively, as pointed out in Refs. \cite{nelson}
and \cite{parker},
one should also see the running behaviour by performing the
rescaling
$g_{\mu\nu}\rightarrow s^{-2} g_{\mu\nu}$ and looking at the
large $s$ limit. Since $\Box\rightarrow s^2\Box$ under this rescaling,
the non-local terms proportional to $\ln\Box$ become relevant
in this limit. From the terms independent of $m$ in Eq.(\ref{eqmov2})
we get
\bea
\alpha (s) &=& \alpha (s=1) -\frac{1}{32 \pi^2} \left(
      (\xi-\frac{1}{6})^2 - \frac{1}{90} \right) \ln s
\label{as}\\
\beta (s) &=& \beta(s=1) -\frac{1}{960 \pi^2} \ln s
\label{bs}
\eea
It is worth noting that the scaling behaviour for $\alpha$ and $\beta$
obtained using
both methods Eqs. (\ref{rge2}-\ref{rge3}) and (\ref{as}-\ref{bs})
are identical.
As far as the Newton constant is concerned, we can obtain
its running behaviour only for $\xi=0$.
In this particular case, the terms proportional to
$m^2$ in Eq. (\ref{eqmov2}) have a logarithmic kernel that
appears in the
combination
\be
-\frac{m^2}{384 \pi^2} \ln(-\frac{\Box}{m^2})
\frac{1}{\Box} (H_{\mu\nu}^{(1)} - 2 H_{\mu\nu}^{(2)})
\label{g-hincha}
\ee
Up to the order we are
working ($O({\cal R}^2)$), the basic tensors
$R_{\mu\nu} - \frac{1}{2} R g_{\mu\nu}$, $H_{\mu\nu}^{(1)}$ and
$H_{\mu\nu}^{(2)}$ are related by
\be
H_{\mu\nu}^{(1)} - 2 H_{\mu\nu}^{(2)} = 4 \Box \left(
R_{\mu\nu} - \frac{1}{2} R g_{\mu\nu} \right)
\label{tensors}
\ee
Therefore, it is natural to express Eq.(\ref{g-hincha})
using this relation, which thus leads to an s-scaling for $G$
\be
G(s) = G(s=1)\left( 1 - \frac{G(s=1) m^2}{6 \pi} \ln{s}
\right )\;\; , (\xi=0)
\ee
that is identical to the $\mu$-scaling Eq. (\ref{Gmu}) for minimal
coupling. For an arbitrary coupling we cannot get the s-scaling
for $G$, since the logarithmic kernel does not appear
in the simple combination Eq.(\ref{g-hincha}).
In this case the relation Eq.(\ref{tensors})
makes the identification of the scaling behaviour
ambiguous.

We shall see in the next Section how to
obtain the running behaviour for
$G$ from the Newtonian potential.

\section{QUANTUM CORRECTIONS TO THE NEWTONIAN POTENTIAL}

As in QED, the vacuum polarization effects contained in $< T_{\mu\nu} >$
induce modifications to the Newtonian potential. We will
now evaluate these
corrections. To begin with, we must obtain the $in-in$
effective equations. To this end, we should express the form
factors as integrals of the massive Euclidean propagator
and replace it by the retarded propagator (see Eq. (\ref{rep})).
However, when computing the Newtonian potential we will consider
only time independent fields. Therefore, $F_{in}(\Box)=F(\nabla^2)$
and the $in-in$ equations are just the Euclidean equations
with $\Box$ substituted by $\nabla^2$.

In the static weak-field approximation we have
\bea
g_{\mu\nu} & = & \eta_{\mu\nu} + h_{\mu\nu} ,
\mid h_{\mu\nu} \mid \ll 1 \nonumber\\
R & = & \frac{1}{2} \nabla^2 h
\eea
where we assumed the Lorentz gauge conditions
$( h^{\mu\nu} - \frac{1}{2} \eta^{\mu\nu} h )_{;\nu} = 0$. For a point
particle with
$T_{\mu\nu} = \delta^{0}_{\mu} \delta^{0}_{\mu\nu} M \delta^{3}({\bf{x}})$,
the trace of the (linearized) Eq. (\ref{eqmov2}) is
\bea
& & \left[
\frac{1}{16 \pi G} \nabla^2 - 2 (3 \alpha + \beta) \nabla^2 \nabla^2
-{1\over 32\pi^2}\left(
\frac{19}{180} - \frac{5 \xi}{3} + 6 \xi^2 -
3(\xi -\frac{1}{6})^{2} \ln(-{\frac{\nabla^2}{\mu^2}})
\right) \nabla^2 \nabla^2  \right.\nonumber\\
& & \left. -{m^2\over 16\pi^2} \left(
\frac{1}{2} (\xi - \frac{1}{6}) (-1 + \ln{\frac{m^2}{\mu^2}}) +
\frac{7}{18}  -
3\xi + 3\xi^2 +
 3 (\xi^2 -\frac{1}{36}) \ln(-{\frac{\nabla^2}{\mu^2}})
\right) \nabla^2
\right] h \nonumber\\
& & + O(m^4) = -M \delta^{3}({\bf{x}})
\label{eqtrace}
\eea
For simplicity we shall
compute only the trace $h$ and not the complete $h_{\mu\nu}$, since this
will simplify the calculations and will be enough for
our purposes. In the
limit $\alpha,\beta \rightarrow 0$, $-h$ is four times
the Newtonian potential.

We shall solve Eq.(\ref{eqtrace}) perturbatively
\be
h = h^{(0)} + h^{(1)}
\ee
The classical contribution $h^{(0)}$ satisfies
\be
( \nabla^{2} - \sigma^{-2} \nabla^{2} \nabla^{2} ) h^{(0)} =
- 16 \pi G M \delta^{3}({\bf{x}})
\label{eqhzero}
\ee
where $\sigma^{-2} =  32 \pi G (3\alpha + \beta)$. The time independent
and spherically symmetric solution is \cite{stelle}
\be
h^{(0)} = \frac{4 G M}{r} (1 - e^{- \sigma r})
\label{hzero}
\ee
The first quantum correction satisfies
\be
( \nabla^{2} - \sigma^{-2} \nabla^{2} \nabla^{2} ) h^{(1)} =
H(\nabla^2) h^{(0)}
\label{h1}
\ee
where
\bea
& & H(\nabla^2) =
{G\over 2\pi}
\left[
\frac{19}{180} - \frac{5 \xi}{3} + 6 \xi^2 -
3 (\xi -\frac{1}{6})^{2}
G(-\frac{\nabla^2}{\mu^2})
\right] \nabla^2 \nabla^2 + {Gm^2\over\pi}\times\nonumber\\
& & \left[
\frac{1}{2} (\xi - \frac{1}{6}) (-1 + \ln{\frac{m^2}{\mu^2}})
+ \frac{7}{18}
- 3
+ 3 \xi^2 +
3 (\xi^2-\frac{1}{36})  G(-\frac{\nabla^2}{m^2}) )
\right] \nabla^2
\label{lineq}
\eea

We now find the solution to this equation. To begin with, we will consider
the limit $\sigma r \rightarrow \infty$, since in this approximation
it is easy to find such a solution. In this limit the classical
potential becomes
\be
h^{(0)} = 4 G M \left(
  \frac{1}{r} + 4 \pi \sigma^{-2} \delta^{3}({\bf{x}}) \right)
\label{h0sr-big}
\ee
Using the action of the kernel $G(-\frac{\nabla^2}{\mu^2})$
on the delta
function
(Eq.(\ref{logdelta})) we find
\bea
& & \frac{1}{64 \pi G^2 M}
 ( \nabla^{2} - \sigma^{-2} \nabla^{2} \nabla^{2} ) h^{(1)} =
A\,\delta^{3}({\bf{x}})
+B\,\nabla^2\delta^{3}({\bf{x}})
+C\,\nabla^2\nabla^2\delta^{3}({\bf{x}})
\nonumber\\
& & + \left[
\frac{3 m^2}{8 \pi^2} (\xi^2 -\frac{1}{36})
     \right] \frac{1}{r^3} +
 \left[
        -\frac{9}{8 \pi^2} (\xi-\frac{1}{6})^2
+\frac{9 \sigma^{-2} m^2}{4 \pi^2} (\xi^2-\frac{1}{36})
     \right] \frac{1}{r^5} +\nonumber\\
& &  \left[
\frac{45 \sigma^{-2}}{2 \pi^2} (\xi-\frac{1}{6})^2
     \right] \frac{1}{r^7}
\label{58}
\eea
where the coefficients $A, B$ and $C$ depend on $m,\mu$ and
$\xi$. The solution to this equation is
\bea
h^{(1)} &=& -\frac{24 G^2 M m^2}{\pi} (\xi^2-\frac{1}{36})
\frac{ \ln{ \frac{r}{r_{\sst 0}} } }{r} -
\frac{12 G^2 M}{\pi} (\xi-\frac{1}{6})^2 \frac{1}{r^3} -
\nonumber\\
& &\frac{72 G^2 \sigma^{-2} M}{\pi}
(\xi-\frac{1}{6})^2 \frac{1}{r^5}
+ O(\sigma^{-4}) + \ldots
\label{h1sr-big}
\eea
The first, second and third terms come from the sources
$r^{-3}, r^{-5}$ and $r^{-7}$.
The dots denote a term proportional to the classical solution
$h^{(0)}$ as well as
corrections
at the origin, which are proportional
to $\delta^3({\bf{x}})$ and its derivatives.
They all come from the sources proportional to $A, B$ and $C$
in Eq.(\ref{58}).
We have not included them
because our quantum corrections are not accurate near the origin. Indeed,
we have derived the modified Einstein equations under the assumptions
$\nabla \nabla {\cal R} \gg {\cal R}^2$ and $m^2 {\cal R} \ll
\nabla \nabla R$. Both conditions
are satisfied for the $\frac{G M}{r}$ potential if $G M \ll r \ll m^{-1}$,
so the origin $r=0$ is excluded.

{}From Eq. (\ref{h1sr-big}) we see that there are two different types of
terms in the quantum correction. The term containing the logarithm
comes from the non-local terms proportional to
$m^2\ln( -\Box)$ in Eq.(\ref{eqmov2}). It
is
qualitatively what we expected from `Wilsonian' arguments.
However, the
coefficient $\frac{24 G^2 M m^2}{\pi} (\xi^2-\frac{1}{36})$ is not exactly
the same as the one derived from the renormalization group equation
(\ref{Gmu}), unless $\xi=0$ or $\xi = \frac{1}{6}$, i.e., minimal or
conformal coupling. This is an important difference with respect to
the QED calculation, and shows that the `Wilsonian' arguments are not
always quantitatively correct. Besides the running of $G$, we have found
additional $r^{-3}$ and $r^{-5}$ corrections.

There are no terms in $h^{(1)}$ that we could associate to a running of
the constants $\alpha$ and $\beta$. This is not surprising because such
a running would imply terms of the form
$\ln{\frac{r}{r_{\sst 0}}} \delta^3({\bf{x}})$, which are ill-defined.
Moreover, we have already
pointed out that our quantum corrections are not valid near
the origin. Therefore, to see the running of these constants we shall
evaluate the exact solution for $h$ and then analyze the limit
$\sigma r \rightarrow 0$ (to this end it is necessary to consider
only the  case $m^2=0$)

As $\sigma$
is proportional to $\vert 3\alpha +\beta\vert^{-{1\over 2}}
l_{Planck}^{-1}$,
this limit makes sense only for very large values of $\alpha$
and $\beta$. Otherwise the limit would apply only for
$r$ smaller than the Planck length $l_{Planck}$, where our
semiclassical calculations are not valid. Experimentally\cite{stelle}
$\vert\alpha\vert ,\vert\beta\vert\leq 10^{19}$, so
$r\ll 10^{11} l_{Planck}$. This is still an extremely
small length. Therefore, what follows should be taken only as an exercise
that shows that it is possible to extract the scale dependence
of $\alpha$ and $\beta$ from the potential $h$.

The calculations for the exact solution to Eq.(\ref{h1}) are
presented in the Appendix. We quote
here the main results. In order to solve the linearized equation of motion
we have to evaluate the action of the kernel
$G(-\frac{\nabla^2}{\mu^2})$ on the Yukawa potential
\be
G(-\frac{\nabla^2}{\mu^2}) \frac{e^{-\sigma r}}{r} =
\ln({\frac{\sigma^2}{\mu^2}}) \frac{e^{-\sigma r}}{r} -
\frac{e^{\sigma r}}{r} Ei(-\sigma r) -
\frac{e^{-\sigma r}}{r} Ei(\sigma r)
\ee
where $Ei(x)$ is the exponential integral function.
With the help of this formula we get the exact solution for
$h^{(1)}$ (see the Appendix) and the limit
$\sigma r \rightarrow 0$ can be
taken. The solution reads
\be
h^{(1)} =
\frac{6 G^2 M \sigma^4 }{\pi} (\xi -\frac{1}{6})^2
\left[
      r\ln(\sigma r)
      + (\gamma-{3\over 2})r \right ] + O(\sigma^6)
\label{h1sr-small}
\ee

It is worth noting that the logarithmic term is exactly
the one expected from
the renormalization group scaling of $3\alpha + \beta$. Indeed, for
small $\sigma r$ the classical potential becomes, up to a constant,
\be
h^{(0)} \simeq - 2 M G \sigma^2 r
\ee
Substituting in the above equation
$G \sigma^2 = (32 \pi (3\alpha+\beta))^{-1}$ by its running
counterpart (Eqs.(\ref{rge2}-\ref{rge3})) with $\mu=\frac{1}{r}$,
one finds
\be
h^{(0)} \simeq\frac{6 G^2 M \sigma^4 }{\pi} (\xi -\frac{1}{6})^2
          r \ln{\frac{r}{r_{\sst 0}}}
\ee
which coincides with the logarithmic term of result
Eq. (\ref{h1sr-small}).

\section{CONCLUSIONS}

Let us summarize the new results contained in this work.
We have obtained the  {\it in-in} effective equations
for an arbitrary gravitational field that include the
backreaction produced by a quantum scalar field of mass $m$.
The equations are non-local, covariant and valid under
the assumption $\nabla\nabla {\cal R}\gg {\cal R}^2$.
In the limit
$m^2\gg -\Box$, the equations become local and reproduce
the Schwinger DeWitt expansion. In the opposite limit,
$m^2\ll -\Box$, the presence of non-local kernels of the
form $\ln (-{\Box\over\mu^2})$ made it possible to read
the scaling behaviour of the gravitational constants
$\alpha$ and $\beta$
under the rescaling of the metric.
This scaling coincides
with the renormalization group predictions.
This is also the case for the $s$-scaling
of the Newton constant, but only for
minimal coupling (this fact has been pointed out in
Ref.\cite{parker}).

Using the {\it in-in} equations we computed the quantum corrections
to the Newtonian potential.
This is our main result.
We have found two types of corrections:
short range corrections that decay faster than ${1\over r}$
and corrections proportional to ${1\over r}\ln {r\over r_0}$,
which we recognized as the scaling of the Newton constant.
This scaling coincides with the renormalization group
prediction only for minimal and conformal coupling.
For other
couplings,
while the
$\mu$-dependence of $G$ is proportional to
$(\xi - {1\over 6})$,
the scaling in the Newtonian potential is  proportional to
$6 (\xi^2 - {1\over 36})$. Therefore, the `Wilsonian' approach
is strictly valid only for $\xi =0$ and $\xi={1\over 6}$.

One of the main motivations behind the present work was the
remark made in Ref.\cite{alamos} about the possibility
of explaining the dark matter problem through the scale
dependence of $G$. In that paper, the running assumed was
the one dictated by the renormalization group equations,
in a theory of gravity containing $R^2$-terms.
{}From our results we see that, in the toy model we have considered,
the renormalization
group behaviour is qualitatively but not quantitatively
reproduced at the level of the
Newtonian potential.
However, at present we can not
draw definite conclusions about the $R^2$ theory, since
we have not included the graviton loop in our calculations.
We hope to clarify this issue in the future.

Finally, we would like to point out that the covariant
effective equations we have found in Section III can also
be used to analyze the effect of scaling in cosmological
situations. For a Robertson Walker metric with scale
factor $a(t)$, we expect local terms of the form
$\ln a^2(t)$ to be contained in the kernel $\ln (-{\Box\over\mu^2})$.
These local terms may have interesting cosmological
and astrophysical consequences, like the generation of a
primordial magnetic field during inflation.\cite{dolgov}
Work in this direction is in progress.

{\it Note added:} While we were writing this article we received
a paper by Donoghue,\cite {donoghue} where the author
calculates the quantum corrections
to the Newtonian potential due to the graviton loop. His results
are qualitatively the same as ours in the case
$m^2=0$. This is to be expected, since
the physical degrees of freedom of the graviton can be treated
as massless scalar fields.

\section{Acknowledgments}

We would like to thank
J.J. Giambiaggi, D.D. Harari and L.E. Oxman for
useful conversations on related matters.
This research was supported by Universidad de Buenos Aires,
Consejo Nacional de Investigaciones Cient\'\i ficas y T\' ecnicas
and by Fundaci\' on Antorchas.

\newpage
\appendix
\section{}

In this Appendix we calculate the first quantum correction $h^{(1)}$
that solves the linearized equation
of motion Eq. (\ref{h1}) for the case $m^2=0$.
The evaluation of the action of the kernel
$G(-\frac{\nabla^2}{\mu^2})$ on the Yukawa potential is accomplished
using Eqs.(\ref{fourier})  and (\ref{gin})
\bea
G(-\frac{\nabla^2}{\mu^2}) \frac{e^{-\sigma r}}{r}  & = &
\int d^3x'\int {d^3k\over (2\pi)^3}
e^{i{\bf k.(x-x')}} \ln {({\bf k}^2)\over\mu^2}\frac{e^{-\sigma r'}}{r'}
 \nonumber\\
& = & \ln{(\frac{\sigma^2}{\mu^2})} \frac{e^{- \sigma r}}{r}
    - \frac{e^{\sigma r}}{r} Ei(-\sigma r)
     - \frac{e^{-\sigma r}}{r} Ei(\sigma r)
\eea
where $Ei(x)$ is the exponential integral function. Taking this
expression into account, the equation of motion reads
\be
\left( \nabla^2 - \sigma^{-2} \nabla^2 \nabla^2 \right) h^{(1)} =
\sum_{i=1}^{5} f_{i} ({\bf{x}})
\ee
where
\bea
f_{1}({\bf{x}}) &=&
        64 G^2 M \sigma^2
\left[
\frac{19}{1440} -\frac{5 \xi}{24} + \frac{3 \xi^2}{4}
        +\frac{3}{8} (\xi-\frac{1}{6})^2 \ln{\mu^2}
        \right] \delta^3({\bf{x}}) \nonumber\\
f_{2}({\bf{x}}) &=&
        64 G^2 M \sigma^2
\left[
\frac{3}{16 \pi} (\xi-\frac{1}{6})^2
        \right] \frac{1}{r^3} \nonumber\\
f_{3}({\bf{x}}) &=&
        64 G^2 M \sigma^4
\left[
        -\frac{19}{5760 \pi} +\frac{5 \xi}{96 \pi} - \frac{3 \xi^2}{16 \pi}
+ \frac{3}{16 \pi} (\xi-\frac{1}{6})^2 \ln{\frac{\sigma}{\mu}}
\right] \frac{e^{-\sigma r}}{r} \nonumber\\
f_{4}({\bf{x}}) &=&
        64 G^2 M \sigma^4
\left[
-\frac{3}{32 \pi} (\xi-\frac{1}{6})^2
        \right] \frac{e^{\sigma r}}{r} Ei(-\sigma r) \nonumber\\
f_{5}({\bf{x}}) &=&
        64 G^2 M \sigma^4
\left[
-\frac{3}{32 \pi} (\xi-\frac{1}{6})^2
        \right] \frac{e^{-\sigma r}}{r} Ei(\sigma r)
\label{fuentes}
\eea

Being a linear equation, we propose a time-independent, spherically
symmetric solution of the form
\be
h^{(1)}({\bf{x}}) = \sum_{i=1}^{5} h^{(1)}_{i}(r)
\ee
where each $h^{(1)}_{i}(r)$ is the solution corresponding to the source
$f_{i}({\bf{x}})$.

Let us denote by ${\cal G}({\bf x}-{\bf x'})$ the Green function
of the operator $\nabla^2-\sigma^{-2}\nabla^2\nabla^2$ (obviously
${\cal G}({\bf x})$ is proportional to $h^{(0)}(\bf x)$).
The solutions $h_i^{(1)}(\bf x)$ are then given by
\be
h_i^{(1)}({\bf x})=\int d^3x' \,\, {\cal G}({\bf x-x'})
f_i({\bf x'})
\ee

The source $f_1(\bf x)$ is proportional to $\delta^3(\bf x)$.
Therefore, $h_1^{(1)}$ is proportional to $h^{(0)}$ and can be
absorbed into the classical parameters.


For $i=2$ we obtain
\bea
h_2^{(1)}(r)&=&{24\over \pi^2 r} G^2M\sigma^2(\xi-{1\over 6})^2
\int_0^{\infty}dt {\ln t\over t(1+t^2)}\sin  (\sigma r t)
\nonumber\\
&=& {6\over \pi r}G^2M\sigma^2(\xi-{1\over 6})^2
\int_0^{\sigma r}dz[e^zE_i(-z)-e^{-z}E_i(z)]
\eea
where the last equality can be proved by taking $r$-derivatives
on both sides and using properties of $E_i(z)$.
Having now the exact first quantum correction, one can analyze the limit
$\sigma r \rightarrow 0$. Using the series expansion for the exponential
integral function \cite{grads}, the quantum correction reduces to
\be
h^{(1)}_2 =
\frac{6 G^2 M \sigma^4 }{\pi} (\xi -\frac{1}{6})^2
\left[
      r\ln(\sigma r)
      + (\gamma-{3\over 2})r + \right ] + O(\sigma^6)
\ee

The other sources can be treated in a similar way. However,
as they are all proportional to $\sigma^4$ (see Eq.(\ref{fuentes})),
the new solutions $h_i^{(1)}\,\,i=3,4,5$ are of order $\sigma^6$.
Therefore,
\be
h^{(1)}({\bf x})=h_2^{(1)}({\bf x}) + O(\sigma^6)
\ee

\end{document}